\documentclass[12pt]{article}
\usepackage[dvips]{graphicx}
\usepackage[usenames, dvipsnames, x11names,rgb]{xcolor}
\usepackage{tikz}
\usetikzlibrary{positioning,shadows,arrows,decorations.pathmorphing,backgrounds,fit,shapes.symbols,chains}
\usepackage{tkz-fct}
\usepackage{pgfplots}
\usepackage{amsfonts}
\usepackage{amsmath}
\usepackage{amssymb}
\usepackage{amsthm}
\usepackage{array,multirow,graphicx}
\usepackage{float}
\usepackage[font=small]{caption}
\usepackage{subcaption}
\usepackage{rotating}
\usepackage{pdflscape}
\usepackage[pdftex]{hyperref}
\hypersetup{colorlinks,%
           citecolor=blue,%
           filecolor=blue,%
           linkcolor=blue,%
           urlcolor=blue,%
           }
\theoremstyle{remark}

\usepackage{lineno}
\pgfplotsset{compat=1.16}
\usepackage{xcolor}
\usepackage{authblk}

\title{Assortativity in cognition}

\author[a]{\normalsize Ennio Bilancini}
\author[b]{\normalsize Leonardo Boncinelli} 
\author[a]{\normalsize Eugenio Vicario}

\affil[a]{\small IMT School for Advanced Studies Lucca, Laboratory for the Analysis of compleX Economic Systems, Piazza S.~Francesco 19, Lucca, 55100, Italy}
\affil[b]{\small University of Florence, Department of Economics and Management, Via delle Pandette 9, 50127 Firenze, Italy}

\begin{document}
\maketitle
\thispagestyle{empty}
\begin{abstract}
    In pairwise interactions assortativity in cognition means that pairs where both decision-makers use the same cognitive process are more likely to occur than what happens under random matching. 
    In this paper we study both the mechanisms determining assortativity in cognition and its effects. In particular, we analyze an applied model where assortativity in cognition helps explain the emergence of cooperation and the degree of prosociality of intuition and deliberation, which are the typical cognitive processes postulated by the dual process theory in psychology.
Our findings rely on agent-based simulations, but analytical results are also obtained in a special case. We conclude with examples showing that assortativity in cognition can have different implications in terms of its societal desirability.
\end{abstract}
\begin{center}
\section*{\small Significance Statement}
\end{center}
\vspace*{-0.25cm}
{\small Assortativity is a phenomenon characterizing social interactions in many contexts and along different dimensions. Our work explores a new dimension of assortativity, occurring at the cognitive level: the agents involved together in an interaction often exhibit similar degrees of cognitive effort. There are theoretical reasons justifying assortativity in cognition, as well as relevant consequences. Assortativity in cognition allows to internalize the external effects of one's own cognition: the partner exerts a similar cognitive effort and hence behaves in a similar way. The evolution of behaviors is significantly affected by assortativity in cognition, with consequences on overall welfare that should be carefully evaluated case by case.}
\clearpage
\pagenumbering{arabic}

\section{Introduction}
Assortativity is a broad concept that can be applied to different contexts. In general, assortativity means that individuals are more likely to be engaged in interactions with people that are similar to them along some dimensions. It is related to homophily: the tendency of individuals to associate and bond with similar others (from Ancient Greek: \emph{homo\^{u}} + \emph{phil\'{i}\={e}}, `love of the same') \cite{currarini2009economic,fu2012evolution}. Assortativity is a widespread phenomenon. A large amount of evidence has been collected showing that individuals often stay and interact with similar others, in some form or another: similarities may refer to belonging to the same cultural group, the same social or ethnic group, or the same religion \cite{mcpherson2001birds}. In network theory, the assortativity coefficient measures the correlation between nodes of similar degree \cite{newman2002assortative}. The effects of assortativity have also been studied extensively, e.g., in genetics \cite{jennings1916numerical,wright1921systems} or for the evolution of cooperation \cite{bergstrom.03,bilancini.boncinelli.wu.18}. If we think of agents as divided in groups according to some characteristic or action, an index of assortativity can be formalized as the difference in probability of matching with an individual of a group conditional on belonging to that same group rather than to a different one \cite{bergstrom.13}.    
Preferences may be used to rationalize different types of assortativity \cite{alger.weibull.13, alger.weibull.16,newton2017preferences}.

The dual process theory is a paradigm that has become prominent in cognitive psychology and social psychology in the last thirty years. In the dual process framework, the decision making is described as an interaction between an intuitive cognitive processes and a deliberative one. Although different approaches emerge from the literature \cite{evans2008dual,kahneman2003perspective,sloman1996empirical}, some common characteristics of the two processes are well established. The intuitive process, also called \textit{system 1} or \textit{type 1}, is fast, automatic, and unconscious, while the deliberative process, also called \textit{system 2} or \textit{type 2}, is slow, effortful and conscious. In evolutionary terms, the intuitive cognitive process is older than the deliberative one, and it is shared with other animals \cite{evans2003two}. The existence of two systems in reasoning and decision making is extended to the domain of learning with associative implicit processes and rule-based explicit processes \cite{reber1989implicit,sun2001implicit}.   

To the best of our knowledge, assortativity in cognition has not been considered and analyzed by the literature so far. In some cases, it is involved or even implied, but the focus was never on it. For instance, priming has been shown to affect the activation of cognitive processes \cite{dijksterhuis1998seeing}, hence interacting partners who are exposed to the same priming are more likely to rely on the same cognitive process. Also, assortativity in actions often implies assortativity in cognition as a byproduct \cite{bear2016intuition}.

%

\section{Sources of assortativity in cognition}
Assortativity in cognition may arise as a consequence of assortativity on other dimensions, such as the characteristics of the interaction or the characteristics of the interacting agents. 

Let $p(D|D)$ be the probability, for a given agent, to interact with a deliberating agent given that the agent is deliberating as well. Following the same notation, $p(D|I)$ is the probability to interact with a deliberating agent given that the agent is deciding intuitively. Let $p(I|I)$ and $p(I|D)$ be defined analogously. There is assortativity in cognition if
\begin{equation}\label{eq:assortativity_1}
    p(D|D)>p(D|I)
\end{equation}
which implies, and is implied by, $p(I|D)<p(I|I)$. 

\subsection{State-based assortativity}
The characteristics of an interaction (e.g., payoffs, information, complexity of choice) vary across interactions but are often the same, or at least similar, for the individuals in the same interaction. When such characteristics determine the likelihood of deliberation, assortativity in cognition emerges. To fix ideas consider a case with two states of the world, $A$ and $B$, that differs in the likelihood that deliberation and intuition are used by agents. State $A$ and state $B$ occur with probabilities $p(A)$ and $p(B)=1-p(A)$, respectively. Agents involved in the same interaction make decisions in the same state. In state $A$ an agent decides intuitively with probability $k_A$ while she deliberates with probability $1-k_A$. Analogously, in state $B$ an agent decides intuitively with probability $k_B$ while she deliberates with probability $1-k_B$. 

In this setting, assortativity in cognition comes out if and only if the likelihood of intuition differs in the two states, i.e., $k_A \neq k_B$ (for the proof see Appendix, Subsection \ref{ass:world}). 

\subsection{Type-based assortativity}
Agents can have heterogeneous characteristics (e.g., skills, abilities, preferences, knowledge) which may determine the likelihood of deliberation. In this case, when the agents participating in the same interaction tend to share the same characteristics, assortativity in cognition emerges. To fix ideas consider the case where the population is composed by two types of agents, $X$ and $Y$, that differ in the likelihood of resorting to deliberation and intuition. The fraction of $X$ agents is equal to $q$ and consequently $1-q$ is the fraction of $Y$ agents. Type $X$ agents and type $Y$ agents decide intuitively with probability $k_X$ and $k_Y$, respectively, while they deliberate with the remaining probability $1-k_X$ and $1-k_Y$. 
Let $p(X|X)$ and $p(X|Y)$ be the probability that a type $X$ interacts with an other type $X$ and with a type $Y$, respectively. There is assortativity in types if $p(X|X)>p(X|Y)$, which implies, and is implied by, $p(Y|X)<p(Y|Y)$.

In this setting, if we assume assortativity in types, then assortativity in cognition comes out if and only if the likelihood of intuition is different for the two types, i.e., $k_X \neq k_Y$ (for the proof see Appendix, Subsection \ref{ass:types}). 

\section{Learning intuitive cooperation through deliberation}\label{section:learning}

Cooperation is a central feature of human behavior that differentiates \textit{Homo Sapiens} from the other species \cite{melis2010human,harari2014sapiens}. When people are cooperative pay a cost to benefit others. The emergence of cooperation as a persistent phenomenon is a major focus of research across different subjects, such as social sciences \cite{bowles2011cooperative} and biology \cite{hamilton1964genetical}. Indeed, the wide empirical evidence on cooperation is puzzling. For social scientists, it is at variance with the paradigmatic rational self-interested individual that is known as \textit{Homo Economicus}, even if other-regarding individuals can have reasons to cooperate \cite{bowles2016moral}. For biologists, competition among individuals is at the basis of natural selection, and this is likely to wipe out cooperators though it is not necessarily the case \cite{koduri2021origin}. In the literature on evolutionary game theory, great attention has been devoted to the mechanisms through which selection can favor the evolution of cooperation \cite{nowak2006five,axelrod1981evolution,rand2013human}. Recently, the cognitive basis of cooperative decision-making has also been explored, both experimentally \cite{rand2012spontaneous,rand2014social,alos2020cognitive} and through theoretical modeling \cite{bear2016intuition,jagau2017general}. 
In the following we show that cognition can play an important role for the evolution of cooperation by the channel of assortativity. By doing so, we exemplify how assortativity in cognition can be incorporated in a fully-fledged model, giving insights on the phenomenon under analysis, namely the emergence of cooperation and the degree of prosociality of intuition and deliberation.

\subsection{The model}
We describe a setting in which agents from a population interact repeatedly in random pairs.
There are two possible types of interaction, the one shot prisoner dilemma, Table \ref{table1}(\subref{oneshot}) that occurs with probability $1-p$, and the repeated interaction, Table \ref{table1}(\subref{repeated}) that occurs with probability $p$. Two actions are available in both interactions, namely cooperation, $C$, and defection, $D$.

\begin{table}[htb]
    \setlength{\extrarowheight}{2pt}

    \begin{subtable}{.5\linewidth}
      \centering
            \begin{tabular}{c|c|c|}
         \multicolumn{1}{c}{} & \multicolumn{1}{c}{$C$}  & \multicolumn{1}{c}{$D$} \\\cline{2-3}
         $C$ & $b$ & $0$ \\\cline{2-3}
       $D$ & $b+c$ & $c$ \\\cline{2-3}
    \end{tabular}
    \caption{\normalfont One shot prisoner dilemma.\label{oneshot}}
    \end{subtable}%
    \begin{subtable}{.5\linewidth}
      \centering
        \begin{tabular}{c|c|c|}

       \multicolumn{1}{c}{} & \multicolumn{1}{c}{$C$}  & \multicolumn{1}{c}{$D$} \\\cline{2-3}
         $C$ &  $b$ & $c$ \\\cline{2-3}
       $D$ & $c$ & $c$ \\\cline{2-3}
    \end{tabular}
    \caption{\normalfont Repeated prisoner dilemma.\label{repeated}}
    \end{subtable}
    \caption{\normalfont Payoff earned by the row player, with the assumption that $b>c>0$.\label{table1}}
\end{table}
\noindent
When the two players in an interaction play $C$, they both earn $b$ irrespectively of the type of interaction. Similarly, when the two players in an interaction play $D$, they both earn $c$ irrespectively of the type of interaction. When the two players choose different actions, the payoffs depend on the type of interaction: in the one-shot prisoner dilemma, the defecting player earns $b+c$ and the cooperating agent earns $0$; in the repeated prisoner dilemma, both players earn $c$. We assume that $b>c>0$, which makes $D$ strictly dominant in the one-shot interaction, and $C$ weakly dominant in the repeated interaction. 
This payoff structure is already used in the literature \cite{bear2016intuition}, with the only difference that $c$ is added in every cell to avoid negative values.


Each agent is able to elaborate the rewards obtained in the past when playing the two different actions, cooperation and defection. This information is stored in the memory of agents. When an agent chooses a certain action then updates the information about the past rewards obtained with that action, keeping unchanged the information about the past rewards obtained with the other action. Indeed the memory of a generic agent $i$ at time $t$, $m_i^t$, is made of two elements, the information about the past rewards obtained in the previous periods when playing cooperation, $\overline{R}_{i,C}^t$, and the information about the past rewards obtained in the previous periods when playing defection, $\overline{R}_{i,D}^t$:

\begin{equation*}
  m_i^t=\{\overline{R}_{i,C}^t, \overline{R}_{i,D}^t  \}
\end{equation*}

In particular, if agent $i$ plays cooperation at time $t$, then the agent's memory is updated in the following way: 

\[ 
\begin{array}{ccl}
    \overline{R}_{i,C}^t & = & (1-\alpha) \overline{R}_{i,C}^{t-1}+\alpha R_i^t\\
    \overline{R}_{i,D}^t & = & \overline{R}_{i,D}^{t-1}
\end{array}
\]
with $\alpha \in \left( 0,1 \right]$ measuring the learning rate and $R_i^t$ being the reward obtained in the last period. Analogously, if agent $i$ plays defection at time $t$, then the agent's memory is updated in the following way:

\[
\begin{array}{ccl}
    \overline{R}_{i,C}^t & = & \overline{R}_{i,C}^{t-1} \\
    \overline{R}_{i,D}^t & = & (1-\alpha) \overline{R}_{i,D}^{t-1}+\alpha R_i^t
\end{array}
\]

This process of memory update, by which we compute a weighted mean between the value stored in memory and the last reward obtained, is a form of reinforcement learning:
it can be seen as myopic \textit{Q-learning} \cite{watkins1992q}, i.e., the case in which agents are not able to make any prediction about the future. 
We note that, when the learning rate $\alpha$ is equal to one, only the last reward obtained for each action matters.


The decision process used by agents relies on either intuition or deliberation, with the latter following a more consequentialist rule (based on best reply) than the former (based on reinforcement learning).
\begin{itemize}
  \item Under intuition the agent is not able to recognize the type of occurring interaction. The intuitive decision is based on the information saved in memory. The action with the highest past reward is chosen: when $\overline{R}_{i,C}^t> \overline{R}_{i,D}^t$ cooperation is chosen, conversely defection is chosen when $\overline{R}_{i,C}^t< \overline{R}_{i,D}^t$. In case of a tie, i.e., when $\overline{R}_{i,C}^t = \overline{R}_{i,D}^t$, each action is chosen with one-half probability.
  \item Under deliberation the agent is able to recognize the type of occurring interaction. The deliberative decision is driven by best-response. Defection is chosen in the one-shot prisoner dilemma because strictly dominant, while cooperation is chosen in the repeated prisoner dilemma, because weakly dominant. In the Appendix (Subsection \ref{append_learning_s2}) we consider a variant in which deliberative decisions are based on myopic Q-learning with finer information, distinguishing between past performance of cooperation and defection under deliberation in the two types of interaction. We show that qualitatively similar results hold in that case as well.
\end{itemize}

We assume that an agent adopts intuition or deliberation depending on the realization of a random variable. In particular, we let $K \in [0,1]$ denote the probability that an agent responds intuitively, so that $1-K$ denotes the probability of deliberation. 
The cognitive processes adopted by two agents interacting together exhibit assortativity, as measured by parameter $A \in [0,1]$. Indeed, with probability $A$ there is a single draw of the random variable, which means that the two agents are forced two use the same cognitive process. With probability $1-A$, there are two independent draws of the random variable, one for each agent, whose cognitive process will be the same or different depending on the realized draws.

We stress that $K$ is homogeneous and exogenous in our model. This is so because our aim is not to study the evolution of dual process reasoning, rather we want to focus on the effects of assortativity in cognition given dual process reasoning, for which the literature has already provided evolutionary arguments \cite{sherry1987evolution,carruthers2006architecture,bear2016intuition}. Quite interestingly, we find that in our model the value of $K$ that maximizes the average payoff is often strictly in between $0$ and $1$ (Appendix, Section \ref{append_optimal_k}).

\subsection{Results}\label{results}
The findings in this subsection are based on simulations over $5000$ periods, with $500$ agents, payoffs $b=4$ and $c=1$, and learning rate $\alpha=0.5$. The code in Python is available at \url{https://github.com/EugenioVicario/Assortativity_in_Cognition}.

\begin{figure}[ht]
  \centering
  \includegraphics[width=1\linewidth]{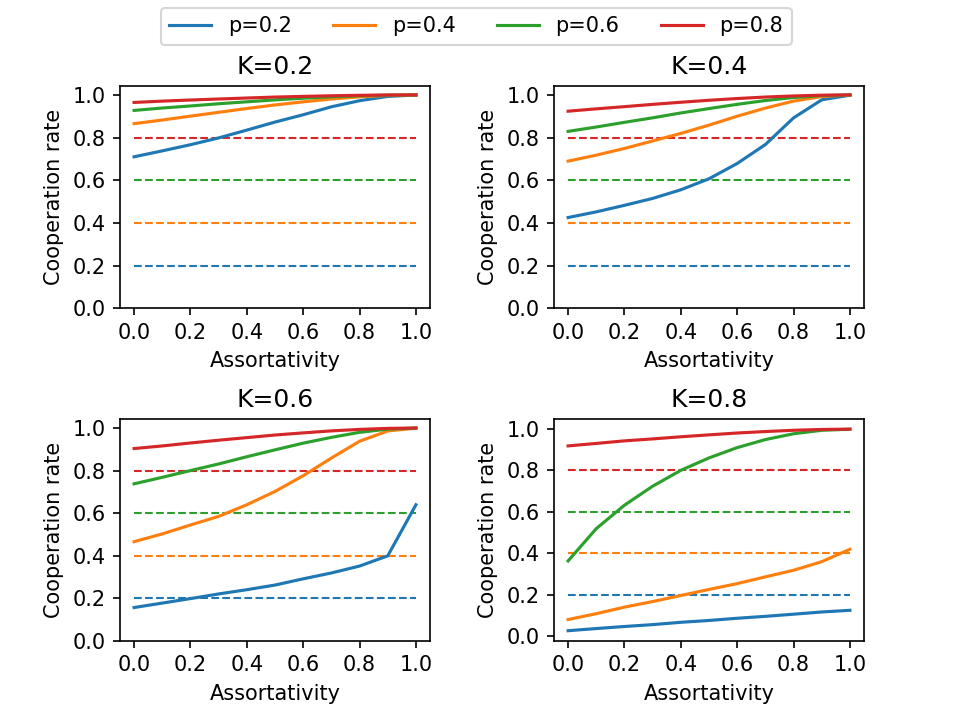}
  \caption{Average cooperation rate varying assortativity in cognition. Each subplot refers to a specific value of $K$. Solid lines represent the average rate of cooperation under intuition, dashed lines represent the average cooperation rate under deliberation, i.e., the value of $p$. Each color refers to a specific value of $p$. }\label{fig:assort_vs_random}
\end{figure}

A first result is that the average cooperation rate increases monotonically in the level of assortativity in cognition. The result is depicted in Figure \ref{fig:assort_vs_random} where solid lines represent the average cooperation rate under intuition as the assortativity in cognition varies. Since the cooperation rate under deliberation is constant and equal to $p$, which depicted with dashed lines in the figure, the result is driven by the increase in cooperation rate under intuition. 
When assortativity in cognition emerges through assortativity in types, it also comes with assortativity in behavior \cite{bear2016intuition}, at least if types are defined including actions. When this is the case, it is impossible to disentangle the effect of assortativity in cognition from the effect of assortativity in behavior. Our result suggests that assortativity in cognition is able to promote cooperation \emph{per sé}, also in the absence of other forms of assortativity.


A second result points to the existing interaction effect between assortativity in cognition and other parameters in the model. In particular, Figure \ref{fig:assort_vs_random} suggests that assortativity in cognition can be a substitute for both the likelihood of repeated interactions, i.e., $p$, and the recourse to deliberation, i.e., $1-K$. Indeed, when $p$ is quite large there is no room for a significant effect of assortativity in cognition, because repeated interactions are frequent and this, in itself, sustains high rates of intuitive cooperation. Also, when $K$ is small every agent frequently deliberates, which implies that often both the agents in an interaction are deliberative, even in the absence of assortativity in cognition.    

A third result is an observation that is independent of assortativity in cognition. The average cooperation rate under intuition, for given $p$ and $A$, increases as $K$ decreases, i.e., the more frequently agents resort to deliberation. Deliberation is able to shape the intuitive heuristic toward cooperation or, in other words, agents learn intuitive cooperation through deliberation.

Finally, a fourth result is about the role of assortativity in cognition in determining whether intuition is more cooperative than deliberation, which is a theme that has been harshly debated in the literature \cite{alos2020cognitive,zaki2013intuitive}. In our model, intuition can be more cooperative than deliberation, or the vice versa can happen, and assortativity in cognition plays a role for this. By looking at Figure \ref{fig:assort_vs_random}, we observe that the average cooperation rate is always higher under intuition than under deliberation when $K$ is quite small or $p$ is quite large. When $K$ is large and $p$ is small, assortativity in cognition matters: indeed, it is often the case that intuition is still more cooperative than deliberation for high values of assortativity, while deliberation turns out to be more cooperative than intuition when assortativity in cognition is small. In this sense, assortativity in cognition helps intuition to be more cooperative than deliberation, in that it enlarges the region in the set of parameters where this holds. 




In the Appendix we provide a robustness check of our results, by considering different entries in the payoff matrix (Subsection \ref{append_payoff_matrix}) and different learning rates (Subsection \ref{append_q-learning}).

\subsection{Markov process}
When the learning rate $\alpha$ is equal to one, the behavior of one agent $i$, given the behavior of all the other agents, in the model can be described through a \textit{discrete-time Markov process} $P$, defined on a finite state space $S$ and characterized by a transition matrix $T$. The state space is made by all the feasible memories of agent $i$, i.e., all the pairs $\{\overline{R}_{i,C}^t, \overline{R}_{i,D}^t  \}$. The transition matrix describes the probabilities of moving from each state to any other. Transition probabilities depend on the current memory, i.e., the state, the parameters $K$ and $p$, and the probability of intuitive cooperation of the rest of the population, denoted by $\overline{x}$. A probability distribution $\pi$ defined on $S$ is a vector of probabilities such that $\sum_{m} \pi_m=1$, where $m \in S$ denotes a memory and $\pi_m$ the probability that the agent has memory $m$. A probability distribution is said invariant if:

\begin{equation*}
    \pi T = \pi
\end{equation*}
In words, an invariant distribution remains unchanged in the Markov process as time progresses. Since the Markov process has a unique recurrent class, the invariant distribution exists and is unique. Once obtained the invariant distribution, the probability of cooperation under intuition for agent $i$ is the sum of probabilities, in the invariant distribution, of states in which $\overline{R}_{i,C}^t > \overline{R}_{i,D}^t$ plus half of the sum of probabilities of states in which $\overline{R}_{i,C}^t = \overline{R}_{i,D}^t$. Indeed, when $\overline{R}_{i,C}^t > \overline{R}_{i,D}^t$ agents cooperate under intuition while they randomly choose the intuitive response in the cases in which $\overline{R}_{i,C}^t = \overline{R}_{i,D}^t$. We denote with $x_i$ the probability of intuitive cooperation in the invariant distribution for agent $i$. Finally, we introduce the consistency condition: in the long run equilibrium of the model, the cooperation rate of agent $i$ is equal to the cooperation rate of the other agents, i.e., $\overline{x}=x_i$. 

For the sake of simplicity, we focus on the case of perfect assortativity. Thus the information about past rewards when playing cooperation $\overline{R}_{i,C}^t $ belongs to $ \{ 0,c,b \}$ and analogously $\overline{R}_{i,D}^t $ belongs to $ \{ c,d \}$, where $d=b+c$. Hence the memory of each agent belongs to the Cartesian product of the column vector $\left[ 0,c,b \right]$ and the row vector $\left[ c,d \right]$.

In the Appendix (Subection \ref{append_markov}) we develop the analysis in detail for the simplifying case of full assortativity, i.e., $A=1$.
\begin{figure}[htb]
  \centering
  \includegraphics[width=0.75\linewidth]{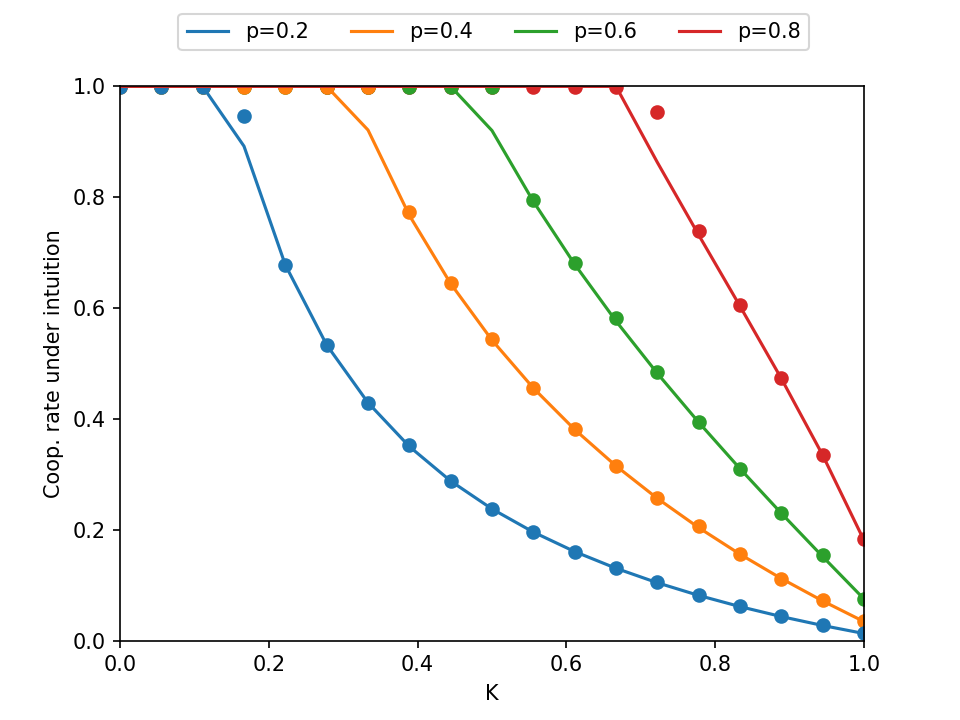}
  \caption{Solid lines are the theoretical frequencies obtained through the long-run Markov chain analysis. Dots are the empirical frequencies obtained through simulations with 500 agents, 5000 time periods, and $A=1$. }\label{fig:simulation_vs_markov}
\end{figure}

Figure \ref{fig:simulation_vs_markov} represents the cooperation rate under intuition, distinguishing between the empirical frequencies obtained through simulations and the theoretical frequencies resulting from the long-run Markov chain analysis. For most values of $p$ and $k$, the theoretical analysis overlap with simulations, with only perceptible differences for cooperation rates that are very close to one. See the Appendix (Subsection \ref{append_analyt_vs_simul}) for more details on this.

\section{Bivalence of assortativity in cognition on payoffs}\label{section:bivalent}
Drawing from the results in Subsection \ref{results}, one may be tempted to conclude that assortativity in cognition is welfare enhancing. In this section, we show that this conclusion would be an overstatement: indeed, the effects of assortativity in cognition on the overall welfare are complicated in general, and hence must be evaluated case by case.

In the previous section we focused on the cooperation rate since the total reward of agents is increasing in it. In the following examples we do not have an action that is always more cooperative than the other action, hence we focus on the average total reward, i.e., the average reward over the whole population along the entire time span.

We replicate the simulations of the previous section changing the types of interaction in which the agents are involved. 
For simplicity, we consider each of the two interactions in Table \ref{table1} combined with a variant of it, in which the two actions are permuted, i.e., the actions have inverted payoff consequences in the two types of interaction. 
In particular, in subsection \ref{subsection:one-shot} we consider two one-shot prisoner dilemmas, while in subsection \ref{subsection:repeated} we consider two repeated prisoner dilemmas. 

\subsection{Double one-shot prisoner dilemma}\label{subsection:one-shot}
Under deliberation agents choose the dominant action, $S$ in game \ref{PD1} and and $F$ in game \ref{PD2}. Let $p$ be the probability of game \ref{PD2}. In this setting, playing the dominated action increases the overall payoff, with the result that miscoordination in behaviors can be beneficial with respect to coordination in the dominant action.
\begin{table}[h]
    \setlength{\extrarowheight}{2pt}

    \begin{subtable}{.5\linewidth}
      \centering
            \begin{tabular}{c|c|c|}
         \multicolumn{1}{c}{} & \multicolumn{1}{c}{$F$}  & \multicolumn{1}{c}{$S$} \\\cline{2-3}
         $F$ & $b$ & $0$ \\\cline{2-3}
       $S$ & $b+c$ & $c$ \\\cline{2-3}
    \end{tabular}
    \caption{\normalfont S dominant action.\label{PD1}}
    \end{subtable}%
    \begin{subtable}{.5\linewidth}
      \centering
        \begin{tabular}{c|c|c|}

       \multicolumn{1}{c}{} & \multicolumn{1}{c}{$F$}  & \multicolumn{1}{c}{$S$} \\\cline{2-3}
         $F$ &  $c$ & $b+c$ \\\cline{2-3}
       $S$ & $0$ & $b$ \\\cline{2-3}
    \end{tabular}
    \caption{\normalfont F dominant action.\label{PD2}}
    \end{subtable}
    \caption{\normalfont $b>c>0$.\label{PDS}}
\end{table}
\begin{figure}[ht]
  \centering
  \includegraphics[width=0.8\linewidth]{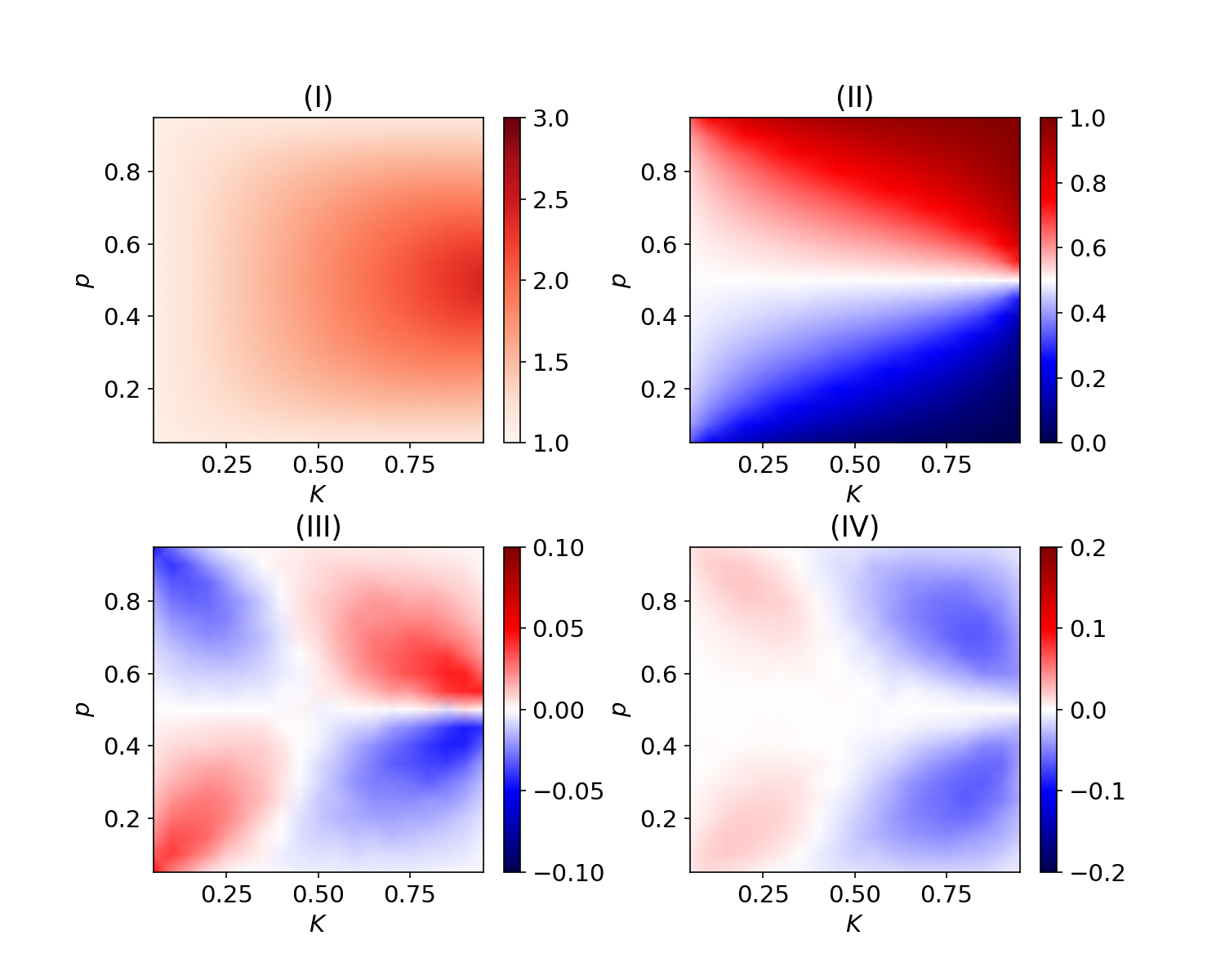}
  \caption{(I) Average reward with $A=0$; (II) Rate of intuitive play of action F when $A=0$; (III) Difference in the rate of intuitive play of action F when $A=1$ and $A=0$; (IV) Average reward with $A=1$ minus average reward with $A=0$. }\label{fig:prisoner_dilemma_assortativity}
\end{figure}

Figure \ref{fig:prisoner_dilemma_assortativity} shows in (IV) that an increase in assortativity is welfare-increasing when $K$ is low and welfare-decreasing when $K$ is high. To grasp the learning effects contributing to this result, we can focus on pairs with one agent intuitive and the other deliberative, given that the main effect of assortativity is to reduce the likelihood of such pairs.
Consider $p>0.5$. As $K$ increases, i.e., agents are more often intuitive, the probability to choose action $F$ gets larger under intuition (Figure \ref{fig:prisoner_dilemma_assortativity}, II).
Suppose first that the intuitive agent chooses $F$. With probability $p$ both agents play $F$, since $F$ is dominant and hence surely chosen by the deliberative agent, yielding no substantial effects on learning. With probability $(1-p)$, the deliberative agent chooses $S$ because dominant, with the result that $S$ performs well and $F$ performs poorly, which makes $S$ more likely to be adopted in the future for both agents.
Suppose now that the intuitive agent chooses $S$. Analogously, with probability $(1-p)$ both agents play $S$, with no substantial effect on learning, while with probability $p$ the deliberative agent chooses $F$ since dominant, which triggers a learning effect. Indeed, in the latter case $F$ performs well and $S$ performs poorly, which makes $F$ more likely to be adopted in the future for both agents.
Please note that $S$ is the welfare-enhancing action, when $p>0.5$.
To complete the reasoning, we observe that the two learning effects described above get weakened when assortativity in cognition increases, due to the reduction in the likelihood that a pair occurs with one agent intuitive and the other deliberative. We note that the former (latter) event is more (less) likely as $K$ increases, because this makes the intuitive player more often choose $F$ (Figure \ref{fig:prisoner_dilemma_assortativity}, II). Therefore, an increase in assortativity reduces the likelihood of playing the dominant action when $K$ is low and increases it when $K$ is high (Figure \ref{fig:prisoner_dilemma_assortativity}, III). Since the dominated action is socially optimal, this leads us to conclude that assortativity in cognition is welfare-enhancing for low values of $K$ and welfare-decreasing for high values of $K$ (Figure \ref{fig:prisoner_dilemma_assortativity}, IV).  

\subsection{Double repeated prisoner dilemma}\label{subsection:repeated}
Under deliberation, agents choose the weakly dominant action, $S$ in game \ref{SH1} and and $F$ in game \ref{SH2}. Let $p$ be the probability of game \ref{SH2}. In this setting average payoffs are maximized when both players choose the weakly dominant action, while other outcomes pay the same.

\begin{table}[h]
    \setlength{\extrarowheight}{2pt}

    \begin{subtable}{.5\linewidth}
      \centering
            \begin{tabular}{c|c|c|}
         \multicolumn{1}{c}{} & \multicolumn{1}{c}{$F$}  & \multicolumn{1}{c}{$S$} \\\cline{2-3}
         $F$ & $c$ & $c$ \\\cline{2-3}
       $S$ & $c$ & $b$ \\\cline{2-3}
    \end{tabular}
    \caption{\normalfont S dominant action.\label{SH1}}
    \end{subtable}%
    \begin{subtable}{.5\linewidth}
      \centering
        \begin{tabular}{c|c|c|}

       \multicolumn{1}{c}{} & \multicolumn{1}{c}{$F$}  & \multicolumn{1}{c}{$S$} \\\cline{2-3}
         $F$ &  $b$ & $c$ \\\cline{2-3}
       $S$ & $c$ & $c$ \\\cline{2-3}
    \end{tabular}
    \caption{\normalfont F dominant action.\label{SH2}}
    \end{subtable}
    \caption{\normalfont $b>c>0$.\label{SHS}}
\end{table}

\begin{figure}[ht]
  \centering
  \includegraphics[width=0.8\linewidth]{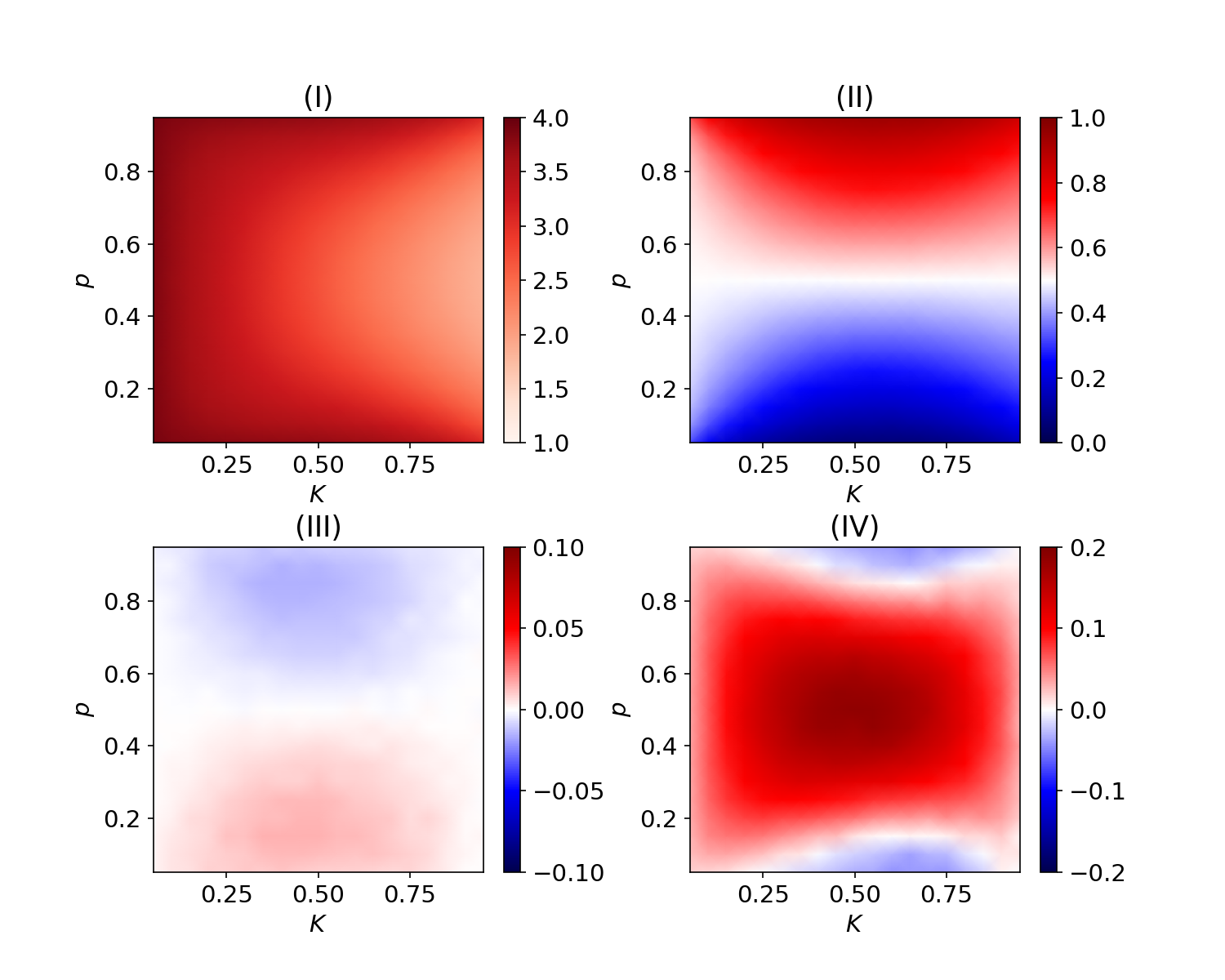}
  \caption{(I) Average reward with $A=0$; (II) Rate of intuitive play of action F when $A=0$; (III) Difference in the rate of intuitive play of action F when $A=1$ and $A=0$; (IV) Average reward with $A=1$ minus average reward with $A=0$. }\label{fig:stag_hunt_assortativity}
\end{figure}

Intuitively, greater deliberation, i.e., a lower $K$, is beneficial because it makes agents choose the weakly dominant action (Figure \ref{fig:stag_hunt_assortativity}, I); the average payoff also increases for extreme values of $p$, close to either $0$ or $1$ (again Figure \ref{fig:stag_hunt_assortativity}, I), because also intuitive agents choose the weakly dominant action most of the time (Figure \ref{fig:stag_hunt_assortativity}, II).

As already pointed out in the previous subsection, assortativity in cognition decreases the probability of interaction between an intuitive agent and a deliberative one, thus increasing the probability of interaction between two intuitive agents and between two deliberative agents. 

On the one hand, an increase of assortativity yields a direct effect on payoffs in that the increased likelihood of two deliberative agents interacting together allows an easier coordination on the weakly dominant action. 

On the other hand, there other effects triggered by learning. To grasp these learning effects, we focus again on pairs with an intuitive agent and a deliberative one. Consider $p>0.5$.
The most likely occurrence here is that agents play game \ref{SH2}, which happens with probability $p$, and that the intuitive agent plays action $F$ (Figure \ref{fig:stag_hunt_assortativity}, II). Since the deliberative agent surely chooses $F$ as well, they obtain the highest payoff $b$, which increases the likelihood of playing action $F$ in the future. The least likely occurrence is that agents play game \ref{SH1}, which happens with probability $1-p$, and that the intuitive agent plays action $S$ (Figure \ref{fig:stag_hunt_assortativity}, II). Since the deliberative agent surely chooses $S$ as well, they obtain the highest payoff $b$, which increases the likelihood of playing action $S$ in the future. Since the action $F$ is the more often the weakly dominant action, given $p>0.5$, the former effect is stronger than the latter. To complete the picture, there are other two cases in which the intuitive agent plays the dominated action, this yielding no substantial effect on learning because both players earn a payoff equal to $c$, even if for different actions. 
Overall, an increase in assortativity in cognition leads to a decrease in the rate at which intuitive agents play the action that is dominant in the interaction occurring with higher probability (Figure \ref{fig:stag_hunt_assortativity}, III). In turn, this has a negative impact on average payoffs, and this impact is greater for extreme values of $p$, close to either $0$ or $1$ (again Figure \ref{fig:stag_hunt_assortativity}, III). It turns out that, for extreme values of $p$, close to either $0$ or $1$, this negative indirect effect through learning more than offsets the positive direct effect on payoffs, resulting in the blue areas in Figure \ref{fig:stag_hunt_assortativity}, IV.

\appendix
\section{Models of assortativity in cognition}
In the following two subsections we provide simple models where assortativity in cognition arises as a consequence of state-based assortativity (\ref{ass:world}) and type-based assortativity (\ref{ass:types}). 

\subsection{State-based assortativity}\label{ass:world}
There are two states of the world, $A$ and $B$, that occur with probabilities $p(A)$ and $p(B)=1-p(A)$, respectively. We assume $p(A)\in (0,1)$. Agents involved together in an interaction are in the same state of the world, i.e., there is full assortativity in the state of the world. Suppose that in state $A$ there is a probability $k_A$ of intuition and a probability $1-k_A$ of deliberation. Analogously, $k_B$ and $1-k_B$ are the probabilities of intuition and deliberation, respectively, in state $B$. We denote with $p(D|D)$ the probability for an agent, conditional on being deliberative, to interact with an agent who is deliberative as well. Following the same notation, $p(D|I)$ is the probability to interact with a deliberative agent, conditional on being intuitive. Assortativity in cognition occurs when:
\begin{equation}\label{eq:assortativity_1}
    p(D|D)>p(D|I)
\end{equation}
From inequality \ref{eq:assortativity_1}, it follows that $p(I|D)<p(I|I)$. Applying Bayes' formula and the definition of conditional probability, inequality \ref{eq:assortativity_1} can be rewritten as:
\begin{equation}\label{eq:assortativity_2}
    \frac{p(D \cap D)}{p(D \cap I)}>\frac{p(D)}{p(I)}
\end{equation}
Let:
\[
\begin{array}{rcl}
    p(D) & = & p(A)(1-k_A)+p(B)(1-k_B) \\
    p(I) & = & p(A)k_A+p(B)k_B \\
    p(D \cap D) & = & p(A)(1-k_A)^2+p(B)(1-k_B)^2 \\     
    p(D \cap I) & = & p(A)(1-k_A)k_A+p(B)(1-k_B)k_B
\end{array}
\]
Substituting in inequality \ref{eq:assortativity_2}, the result is:
\begin{equation}
    \frac{p(A)(1-k_A)^2+p(B)(1-k_B)^2}{p(A)(1-k_A)k_A+p(B)(1-k_B)k_B} > \frac{p(A)(1-k_A)+p(B)(1-k_B)}{p(A)k_A+p(B)k_B}
\end{equation}
Multiplying for the inverse of the right hand side, we obtain:
\begin{equation}\label{eq:lunga}
    \frac{p(A)^2(1-k_A)^2k_A + p(B)^2(1-k_B)^2k_B + p(A)p(B)\left[ (1-k_A)^2k_B + (1-k_B)^2k_A \right]}{p(A)^2(1-k_A)^2k_A + p(B)^2(1-k_B)^2k_B + p(A)p(B)\left[ (1-k_A)(1-k_B)(k_A+k_B) \right]}>1
\end{equation}
The numerator and denominator are identical except for the two parts inside square brackets. Then inequality \ref{eq:lunga} can be rewritten as:
\begin{equation}
    (1-k_A)^2k_B + (1-k_B)^2k_A> (1-k_A)(1-k_B)(k_A+k_B)
\end{equation}
The inequality can be reduced to:
\begin{equation}\label{eq:finale_world}
    (k_A-k_B)^2>0
\end{equation}
In conclusion, there exists assortativity in cognition, as defined in inequality \ref{eq:assortativity_1}, if and only if $k_A \neq k_B$.

\subsection{Type-based assortativity}\label{ass:types}
The population comprises two types of agents, $X$ and $Y$. The fraction of $X$ agents is equal to $p(X)$ and, consequently, $p(Y)=1-p(X)$ is the fraction of $Y$ agents. We assume $p(X)\in (0,1)$. The two types differ in the probability to respond intuitively or deliberately. Assume that type $X$ agents respond intuitively with probability $k_X$ while they deliberate with probability $1-k_X$. The probability of intuitive and deliberative answers for type $Y$ agents are, respectively, $k_Y$ and $1-k_Y$. Assume that there exists assortativity in types, i.e., the probability to interact with a type $X$ agent is greater for a type $X$ agent than for a type $Y$ agent:
\begin{equation}\label{eq:ass_type}
    p(X|X)>p(X|Y)
\end{equation}
From inequality \ref{eq:ass_type}, it follows $p(X|Y)<p(Y|Y)$. As in the previous subsection, assortativity in cognition is defined as:
\begin{equation}
    p(D|D)>p(D|I)
\end{equation}
Applying again Bayes’ formula and the definition of conditional probability, let::
\begin{equation}\label{eq:ass_repeat}
    \frac{p(D \cap D)}{p(D \cap I)}>\frac{p(D)}{p(I)}
\end{equation}
\[
\begin{array}{rcl}
    p(D)  & = & p(X)(1-k_X)+p(Y)(1-k_Y)  \\
    p(I)  & = & p(X)k_X+p(Y)k_Y \\
    p(D \cap D)  & = & p(X)p(X|X)(1-k_X)^2 + p(X)p(Y|X)(1-k_X)(1-k_Y) + \\
                 &   & + p(Y)p(X|Y)(1-k_Y)(1-k_X) + p(Y)p(Y|Y)(1-k_Y)^2  \\
    p(D \cap I)  & = & p(X)p(X|X)(1-k_X)k_X + p(X)p(Y|X)(1-k_Y)K_X +  \\
                 &   & + p(Y)p(X|Y)(1-k_X)k_Y + p(Y)p(Y|Y)81-k_Y)k_Y 
\end{array}
\]
Proceeding analogously as for state-based assortativity, if $p(X)\neq 0$, $p(Y)\neq 0$ and $k_X \neq k_Y$, then inequality \ref{eq:ass_repeat} is equivalent to:
\begin{equation}
    p(D|D)>p(D|I)
\end{equation}
In conclusion, if there is assortativity in types and the probability of deliberation is different for the two types, then assortativity in cognition emerges.

\section{Theoretical analysis}
\subsection{Markov chain}\label{append_markov}

To describe the transition probability from state to state, we firstly need to write the probabilities of obtaining a certain reward with cooperation and defection conditioned on the state:
\[
\begin{array}{lcl}
    P \left[ R_C=r_c | m \right] & &\text{with } r_c \in \left[ 0,c,b \right] \text{ and } m \in S  \\
    P \left[ R_D=r_d | m \right] & &\text{with } r_d \in \left[ c,d \right] \text{ and } m \in S
\end{array}
\]

The probabilities of obtaining a certain reward with cooperation and defection depend on the state on which they are conditioned. The states space can be partitioned in three subsets. In the first one, labeled with $S_1$, there are the states in which $\overline{R}_{i,C}^t > \overline{R}_{i,D}^t$. In the second group $S_2$ there are the states in which $\overline{R}_{i,C}^t = \overline{R}_{i,D}^t$, and in the third group $S_3$ the remaining states in which $\overline{R}_{i,C}^t < \overline{R}_{i,D}^t$. The probabilities of obtaining a certain reward with cooperation and defection depend on the parameters $K$, $p$, and $\overline{x}$.

In $S_1$ the probabilities of different rewards are:
\[
\begin{array}{ccl}
   P \left[ R_C=0 | m \in S_1 \right]  & = & K(1-p)(1-\overline{x})  \\
   P \left[ R_C=c | m \in S_1 \right]  & = & Kp(1-\overline{x})  \\
   P \left[ R_C=b | m \in S_1 \right]  & = & (1-K)p+K\overline{x}  \\
   P \left[ R_D=c | m \in S_1 \right]  & = & (1-K)(1-p)  \\
   P \left[ R_D=d | m \in S_1 \right]  & = & 0  
\end{array}
\]

In $S_2$ the probabilities of different rewards are:
\[
\begin{array}{ccl}
   P \left[ R_C=0 | m \in S_2 \right]  & = & \frac{1}{2} K(1-p)(1-\overline{x})  \\
   P \left[ R_C=c | m \in S_2 \right]  & = & \frac{1}{2} Kp(1-\overline{x})  \\
   P \left[ R_C=b | m \in S_2 \right]  & = & (1-K)p+\frac{1}{2} K\overline{x}  \\
   P \left[ R_D=c | m \in S_2 \right]  & = & (1-K)(1-p) +\frac{1}{2} K(1-\overline{x})+\frac{1}{2}Kp\overline{x}  \\
   P \left[ R_D=d | m \in S_2 \right]  & = & \frac{1}{2} K(1-p)\overline{x}  
\end{array}
\]

In $S_3$ the probabilities of different rewards are:
\[
\begin{array}{ccl}
   P \left[ R_C=0 | m \in S_3 \right]  & = & 0  \\
   P \left[ R_C=c | m \in S_3 \right]  & = & 0  \\
   P \left[ R_C=b | m \in S_3 \right]  & = & (1-K)p  \\
   P \left[ R_D=c | m \in S_3 \right]  & = & (1-K)(1-p)+K(1-\overline{x})+Kp\overline{x}  \\
   P \left[ R_D=d | m \in S_3 \right]  & = & K(1-p)\overline{x}  
\end{array}
\]

\begin{figure}
    \centering

\begin{tikzpicture}

  \node (1) at (-1,3) [shape=rectangle, draw,fill=red!50, minimum width=2cm, minimum height = 1cm] {$m=\{ 0,c \}$};
  
  \node (2) at (5,3) [shape=rectangle, draw,fill=violet!50, minimum width=2cm, minimum height = 1cm] {$m=\{ c,c \}$};
  
  \node (3) at (11,3) [shape=rectangle, draw,fill=cyan!50, minimum width=2cm, minimum height = 1cm] {$m=\{ b,c \}$};
  
  \node (4) at (-1,0) [shape=rectangle, draw,fill=red!50, minimum width=2cm, minimum height = 1cm] {$m=\{ 0,d \}$};
  
  \node (5) at (5,0) [shape=rectangle, draw,fill=red!50, minimum width=2cm, minimum height = 1cm] {$m=\{ c,d \}$};
  
  \node (6) at (11,0) [shape=rectangle, draw,fill=red!50, minimum width=2cm, minimum height = 1cm] {$m=\{ b,d \}$};
  
  \node (7) at (4.2,4)[draw=none]{};
  \node (8) at (5.8,4)[draw=none]{};
  
  \draw[-stealth]  (2)-- (1)node[midway,above]{$P \left[ R_C=0 | \{c,c\} \right]$};
  \draw[-stealth]  (2)-- (3)node[midway,above]{$P \left[ R_C=b | \{c,c\} \right]$};
  \draw[-stealth]  (2)-- (4)node[midway,above]{$0$};
  \draw[-stealth]  (2)-- (5)node[pos=.8,right]{$P \left[ R_D=d | \{c,c\} \right]$};
  \draw[-stealth]  (2)-- (6)node[midway,above]{$0$};
  \draw (4.2,3.5)--(7.center);
  \draw (7.center)--(8.center)node[midway,above]{$P \left[ R_C=c | \{c,c\} \right] + P \left[ R_D=c | \{c,c\} \right]$};
  \draw[-stealth] (8.center)--(5.8,3.5);
\end{tikzpicture}
     \caption{Transition probabilities from state $\{c,c\}$. Colors blue, violet, and red denote states belonging to $S_1$, $S_2$, and $S_3$, respectively. \label{fig:transition_diagram}}
    
\end{figure}
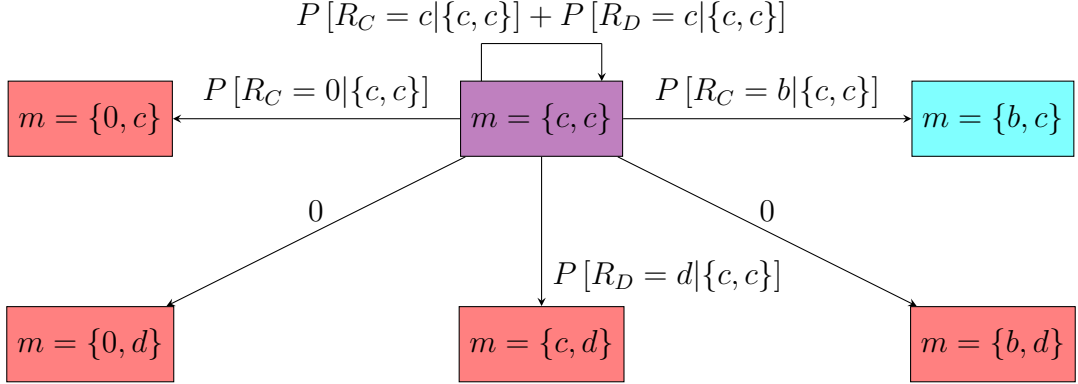

Starting from the probabilities of obtaining a certain reward with cooperation and defection conditioned on the memory, it is straightforward to build the transition probabilities between different states. To give an example, we focus on the transition probabilities from the state $\{ c,c \}$. The probability of transition in one step from $\{ c,c \}$ to $\{ 0,d \}$ or $\{ b,d \}$ is equal to zero. Indeed, at least two steps are required to change both the rewards stored in memory. The probability of the transition from $\{ c,c \}$ to $\{ c,d \}$ is equal to $P \left[ R_D=d | \{c,c\} \right]$, while the probabilities of the transitions from $\{ c,c \}$ to $\{ 0,c \}$ and $\{ b,c \}$ are, respectively, $P \left[ R_C=0 | \{c,c\} \right]$ and $P \left[ R_C=b | \{c,c\} \right]$. Finally, the probability to remain in $\{ c,c \}$ is equal to the probability of obtaining again $c$ from cooperation plus the probability of obtaining again $c$ from deliberation: $P \left[ R_C=c | \{c,c\} \right] + P \left[ R_D=c | \{c,c\} \right]$. See Figure \ref{fig:transition_diagram} for graphical representation.

In the transition matrix the entry $T_{m_im_j}$ represents the probability to have a transition from state $m_i$ to state $m_j$. The summation of all the entries along every row is equal to one. 
\[
 \bordermatrix{ & \{0,c \} & \{c,c \} & \{b,c \} & \{0,d \} & \{c,d \} & \{b,d \} \cr
          \{0,c \} &T_{\{0,c\}\{0,c\}}& 0&T_{\{0,c\}\{b,c\}}&  T_{\{0,c\}\{0,d\}}& 0& 0\cr
          \{c,c \} &T_{\{c,c\}\{0,c\}}&  T_{\{c,c\}\{c,c\}} & T_{\{c,c\}\{b,c\}}& 0& T_{\{c,c\}\{c,d\}}& 0\cr
          \{b,c \} & T_{\{b,c\}\{0,c\}}&  T_{\{b,c\}\{c,c\}}& T_{\{b,c\}\{b,c\}}& 0& 0& 0\cr
          \{0,d \} & T_{\{0,d\}\{0,c\}}& 0&0&  T_{\{0,d\}\{0,d\}}& 0&  T_{\{0,d\}\{b,d\}}\cr
          \{c,d \} &0&  T_{\{c,d\}\{c,c\}}&0& 0&   T_{\{c,d\}\{c,d\}}& T_{\{c,d\}\{b,d\}}\cr
          \{b,d \} &0 &0 & T_{\{b,d\}\{b,c\}}& 0& 0& T_{\{b,d\}\{b,d\}}} \qquad
\]


As stated above the Markov chain has a unique recurrent class and thus the invariant distribution exists and is unique:

\begin{equation*}
    \pi T = \pi
\end{equation*}
The probability of each state in $\pi$ is a function of the probability of intuitive cooperation $\overline{x}$. 

\begin{equation*}
    \pi = \left[ \pi_{\{0,c\}}(\overline{x}),\pi_{\{c,c\}}(\overline{x}),\pi_{\{b,c\}}(\overline{x}),\pi_{\{0,d\}}(\overline{x}),\pi_{\{c,d\}}(\overline{x}),\pi_{\{b,d\}}(\overline{x})  \right]
\end{equation*}
The probability of intuitive cooperation is equal to the probability of being in a state belonging to $S_1$ plus half the probability of being in a state belonging to $S_2$.

\begin{figure}[htb]
    \centering
    \includegraphics[width=0.7\linewidth]{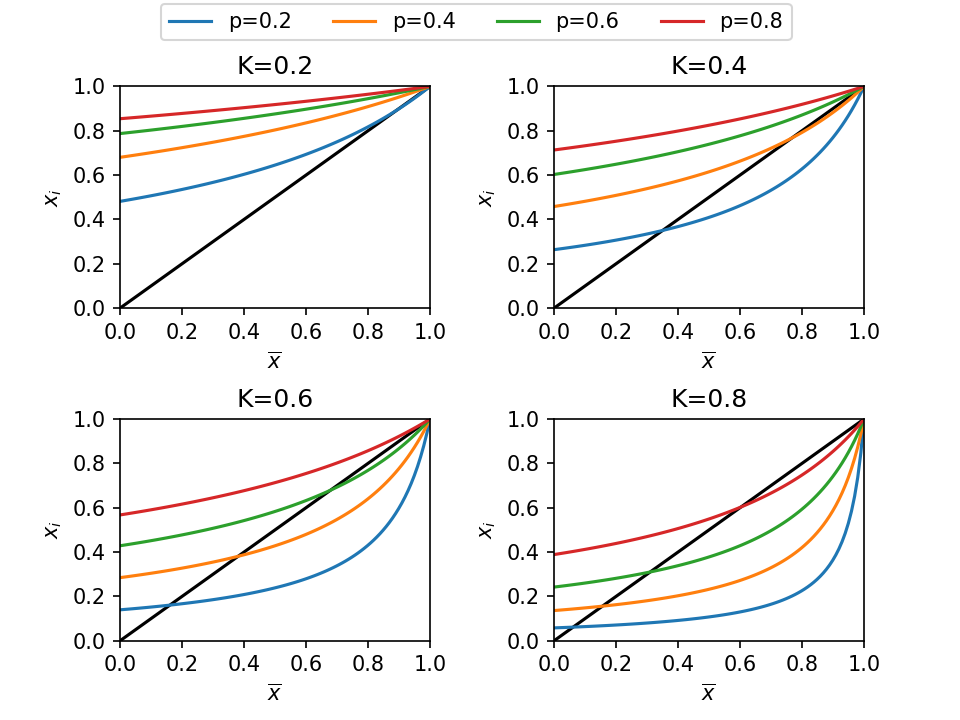}
    \caption{Colored lines represent $x_i(\overline{x})$ for $A=1$ and different values of $p$. The intersection with the $45^{\circ}$ line identifies the equilibria satisfying the consistency condition $x_i = \overline{x}$. Each subplot refers to a different value of $K$.}
    \label{fig:markov_x}
\end{figure}

\begin{equation}\label{intuitive_cooperation}
    x_i=\pi_{\{b,c\}}(\overline{x})+\frac{1}{2}\pi_{\{c,c\}}(\overline{x})
\end{equation}

For the consistency condition we should have:

\begin{equation}\label{consistency_condition}
    x_i = \overline{x}
\end{equation}



Solving Equation \ref{intuitive_cooperation}, considering the consistency condition in equation \ref{consistency_condition}, we obtain the equilibrium values of $x_i$ and $\overline{x}$, let $x=x_i=\overline{x}$. For all the values of $K\in (0,1)$ and $p\in (0,1)$, $x=1$ is always a solution. When all the agents intuitively cooperate, the reward of cooperation remains higher than the reward of defection whatever the type of interaction and the cognitive mode chosen. When large values of $K$ are associated with low values of $p$, a second solution emerges in between $0$ and $1$. 



The analysis carried out so far is for given aggregate behavior, focusing on the values of $x$ where the resulting individual behavior is consistent, i.e. coincides, with the aggregate behavior. When this does not happen, it is natural to ask how the aggregate behavior evolves in response to individual behaviors disconfirming it. We posit that the aggregate behavior will decrease over time when $x_i < \overline{x}$, while it will increase when $x_i > \overline{x}$. Avoiding the burden of introducing a formal dynamical model, we rely on this assumption and on the observation from Figure \ref{fig:markov_x} on the shapes of $x_i(\overline{x})$, to conclude that: when only the $x=1$ solution exists it is an attractor, while, when also another solution exists, this latter solution is attractive and $x=1$ is no longer so.

\subsection{Analytical vs.~simulative results}\label{append_analyt_vs_simul}
\begin{figure}[h]
    \centering
    \includegraphics[width=1\linewidth]{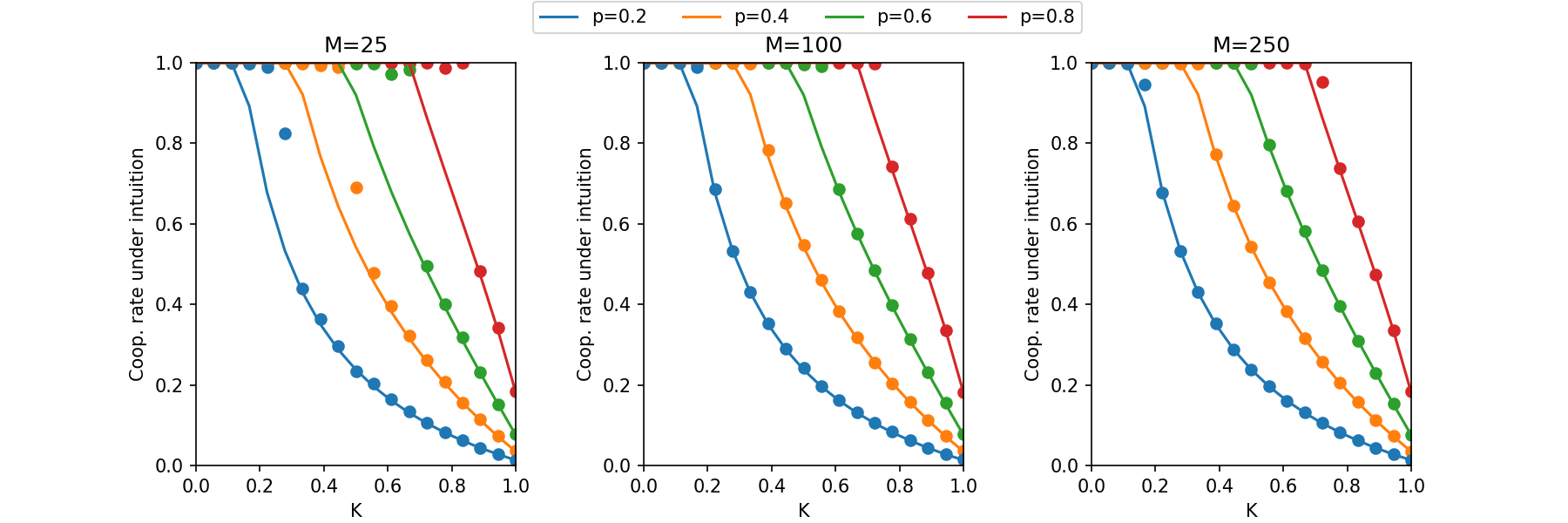}
    \caption{Three different values of the population size $M$. Solid lines represent the theoretical frequencies obtained through the long-run Markov chain analysis. Dots represent the empirical frequencies obtained through simulations with $500$ agents, $5000$ time periods, and $A=1$.}
    \label{fig:change_M}
\end{figure}

In Figure \ref{fig:change_M} we observe that there is a discrepancy between analytical and simulative results. In fact, simulations perfectly predict analytical results when cooperation rate under intuition is equal to one and when it is small enough. When cooperation rate under intuition is close to $1$, but different from $1$, simulations tend to overestimate it.  

Overestimation occurs with perfect assortativity when, at some point $t$ in time, we have that $R_C^{t}>R_D^{t}$ for all agents. In the following periods, all the agents always cooperate under intuition. The system reaches the equilibrium with $x=1$ and, unless perturbations are introduced, the system can not leave such equilibrium. When two equilibria exist and the the minimum $x$ in those equilibria is close to $1$, it is possible that, during the process of convergence to the equilibrium with $x<1$ (which is the attractor), the dynamic reaches the equilibrium with $x=1$ due to stochastic realizations of intuitive behavior as cooperation, which is quite likely since $x$ is close to $1$. In Figure \ref{fig:change_M} we show that the greater is the number of agents the lower is the overestimation, because having that all realized behaviors are cooperative becomes less likely.

\section{Robustness analysis}
\subsection{Payoff matrix}\label{append_payoff_matrix}
\begin{figure}[h]
    \centering
    \includegraphics[width=0.9\linewidth]{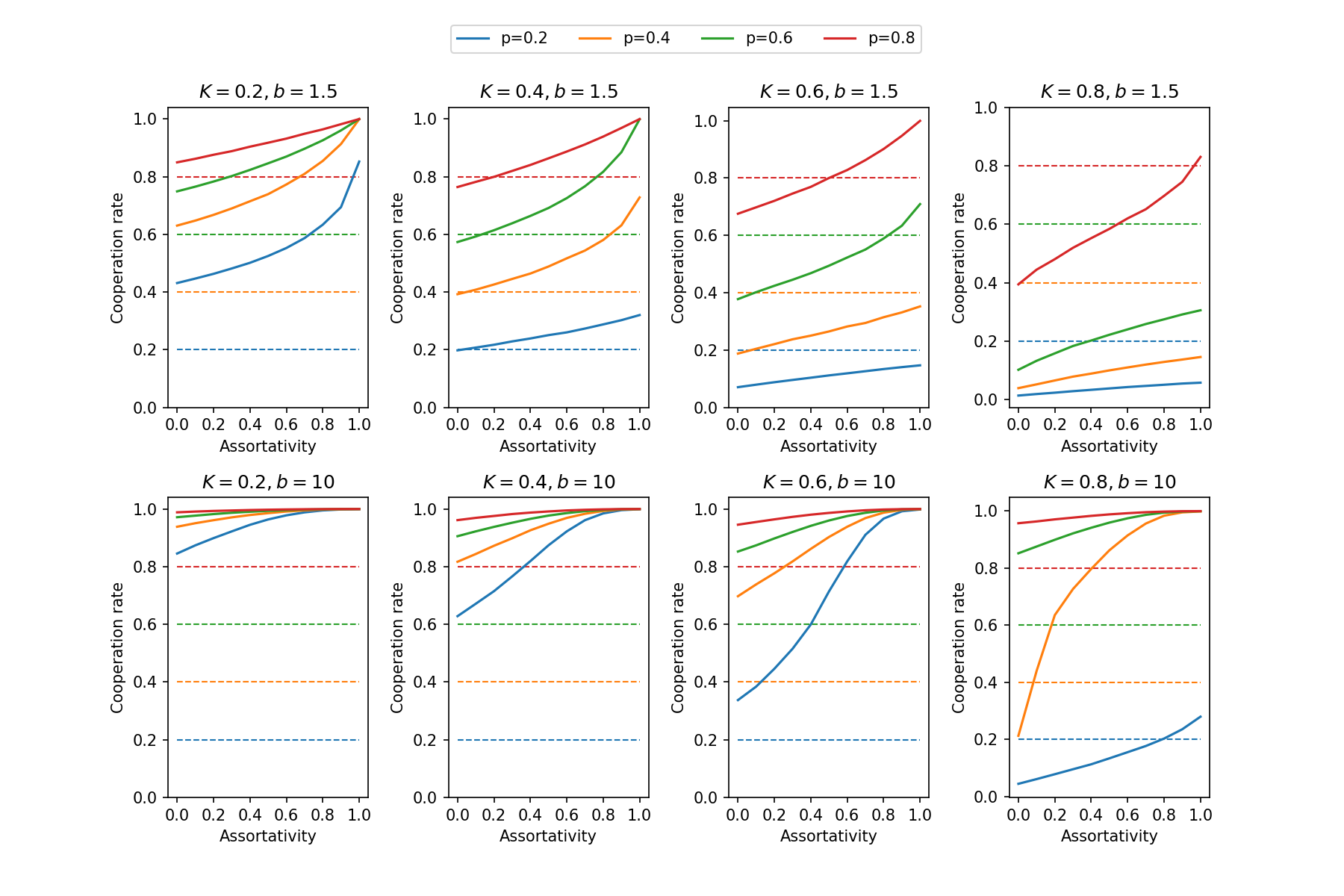}
    \caption{Average cooperation rate varying assortativity in cognition. Each subplot refers to a specific value of $K$ and $b$. Solid lines represent the average rate of cooperation under intuition, dashed lines represent the average cooperation rate under deliberation, i.e., the value of $p$. Each color refers to a specific value of $p$.}
    \label{fig:robust_payoff_s1}
\end{figure}

The payoff matrices of the two types of interaction are equivalent to the games described by \cite{bear2016intuition}, in that they are obtained from theirs by adding $c$ in every entry to avoid negative payoffs.

\begin{table}[hb]
    \setlength{\extrarowheight}{2pt}

    \begin{subtable}{.5\linewidth}
      \centering
            \begin{tabular}{c|c|c|}
         \multicolumn{1}{c}{} & \multicolumn{1}{c}{$C$}  & \multicolumn{1}{c}{$D$} \\\cline{2-3}
         $C$ & $b$ & $0$ \\\cline{2-3}
       $D$ & $b+c$ & $c$ \\\cline{2-3}
    \end{tabular}
    \caption{One shot prisoner dilemma.\label{oneshot}}
    \end{subtable}%
    \begin{subtable}{.5\linewidth}
      \centering
        \begin{tabular}{c|c|c|}

       \multicolumn{1}{c}{} & \multicolumn{1}{c}{$C$}  & \multicolumn{1}{c}{$D$} \\\cline{2-3}
         $C$ &  $b$ & $c$ \\\cline{2-3}
       $D$ & $c$ & $c$ \\\cline{2-3}
    \end{tabular}
    \caption{Repeated prisoner dilemma.\label{repeated}}
    \end{subtable}
\end{table}

In the simulations presented in the paper we consider $b=4$ and $c=1$ as in \cite{bear2016intuition}. Results are qualitatively similar if we consider different values of $b$ and $c$, indeed cooperation rates under intuition monotonically increase as assortativity in cognition increases. Moreover, the lower $p$, the lower the cooperation rate and also the greater $K$, the lower the cooperation rate. Furthermore, we notice that the greater is $b$, maintaining $c$ constant, the greater is the rate of cooperation.


\subsection{Q-Learning}\label{append_q-learning}

\begin{figure}[htb]
    \centering
    \includegraphics[width=0.75\linewidth]{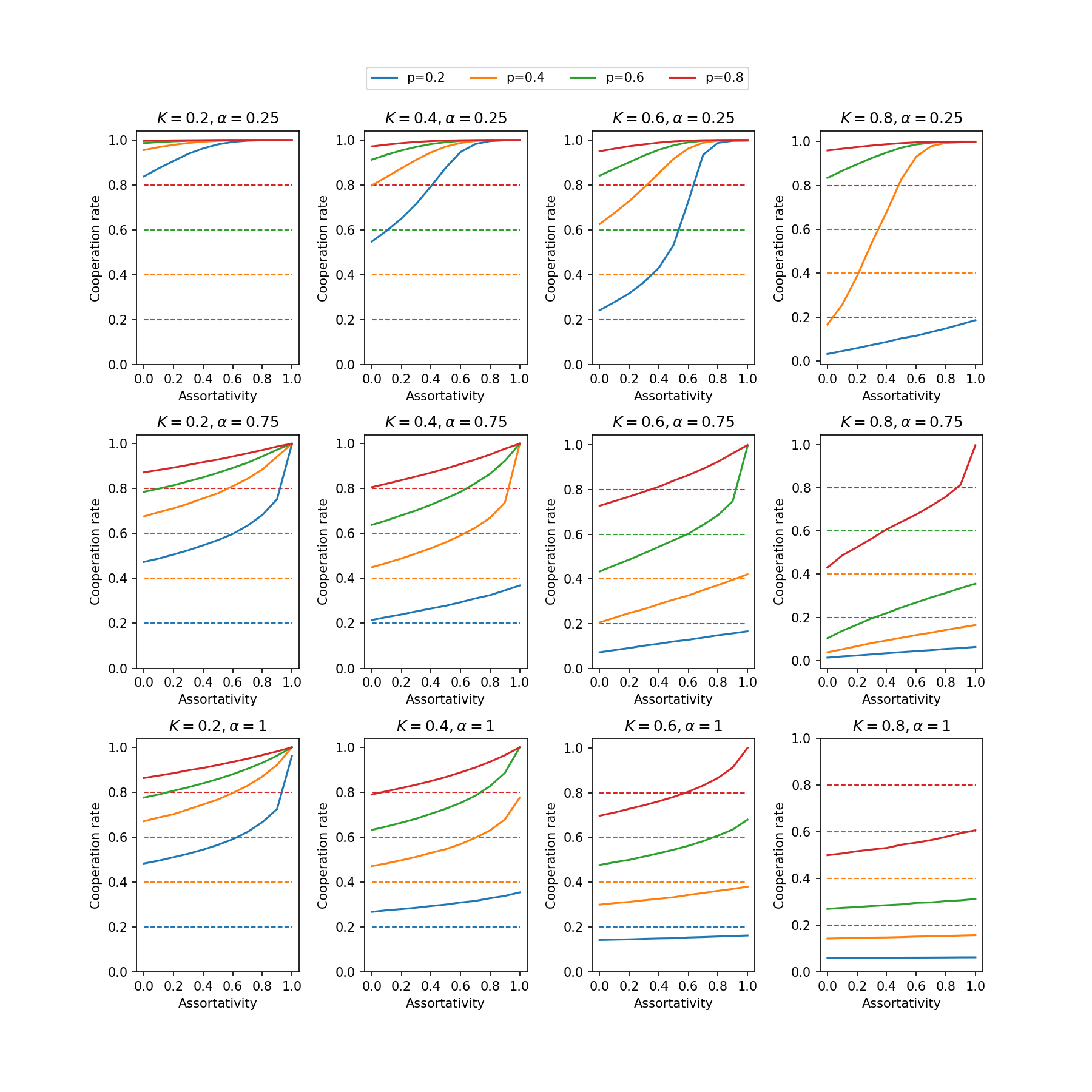}
    \caption{Average cooperation rate for different levels of assortativity in cognition. Each subplot refers to a specific value of $K$ and $\alpha$. Solid lines represent the average rate of cooperation under intuition, dashed lines represent the average cooperation rate under deliberation (which coincide with $p$). Each color refers to a specific value of $p$.}
    \label{fig:Q-learning}
\end{figure}
The standard formulation of Q-learning update is
\begin{equation}\label{eq:q-learning}
    Q_{t}(x,a)=(1-\alpha_{t})Q_{t-1}+\alpha_{t} \left[ r_{t} + \gamma V_{t-1}(y_{t}) \right]
\end{equation}
where $x$ is the state and $a$ is the action performed, $\alpha_t$ is the learning rate, $r_{t}$ is the reward obtained in period $t$, and $V_{t-1}$ is the future value of state $y_t$ that can be reached playing action $a$ in state $x$, which is discounted by the parameter $\gamma$ \cite{watkins1992q}. In our formulation, $\gamma$ is equal to zero, representing the situation in which agents are myopic, i.e., they are unable to make any prediction about future rewards.

Since our results in the main text are given for $\alpha=0.5$, here we explore their robustness by considering different learning rates. Figure \ref{fig:Q-learning} shows the cooperation rate attained under intuition for $\alpha$ taking values $0.25$, $0.75$ and $1$, as $K$, $p$, and $A$ change. In particular, $K$ and $p$ range from $0.2$ to $0.8$ with steps of $0.2$, while instead $A$ ranges from $0$ to $1$ with steps of $0.1$. We notice that the quality of results does not vary as $\alpha$ changes, indeed cooperation rates under intuition monotonically increase as assortativity in cognition increases. Moreover, the lower $p$, the lower the cooperation rate and also the greater $K$, the lower the cooperation rate. More precisely, we observe that a larger weight given to the past information, i.e., a smaller $\alpha$, leads to greater cooperation under intuition and, hence, overall.

\subsection{Q-Learning under deliberation}\label{append_learning_s2}

In this subsection we introduce a learning process for deliberation as well. Agents are characterized by three different memories. The first is the memory through which agents take an intuitive decision, $m_i^t$, that comprises two elements. One is the information about the past rewards obtained in the previous periods when playing cooperation, $\overline{R}_{i,C}^t$, and the other is information about the past rewards obtained in the previous periods when playing defection, $\overline{R}_{i,D}^t$:

\begin{equation*}
  m_i^t=\{\overline{R}_{i,C}^t, \overline{R}_{i,D}^t  \}
\end{equation*}

The second and the third memories are instead used to take decisions under deliberation, one for each of the two different games. The second memory, $m_{i,0}^t$, comprises two elements: one is $\overline{R}_{i,C,0}^t$, which is a statistics of the payoffs obtained in the past when agent $i$ cooperates under deliberation in the repeated prisoner dilemma; the other is $\overline{R}_{i,D,0}^t$, which is a statistics of the payoffs obtained in the past when agent $i$ defects under deliberation in the repeated prisoner dilemma. Agent $i$ at time $t$, when choosing under deliberation in the repeated prisoner dilemma, takes the action associated to the highest between $\overline{R}_{i,D,0}^t$ and $\overline{R}_{i,C,0}^t$. The third memory is $m_{i,1}^t$, and it similarly comprises two elements: $\overline{R}_{i,C,1}^t$ and $\overline{R}_{i,D,1}^t$, which are statistics of past payoffs, for cooperation and defection, respectively,  when agent $i$ has deliberated in the one shot prisoner dilemma. Both the memories used for deliberation, $m_{i,0}^t$ and $m_{i,1}^t$, are updated following the same procedure as $m_i^t$. 

\begin{figure}[hbt]
    \centering
    \includegraphics[width=1\linewidth]{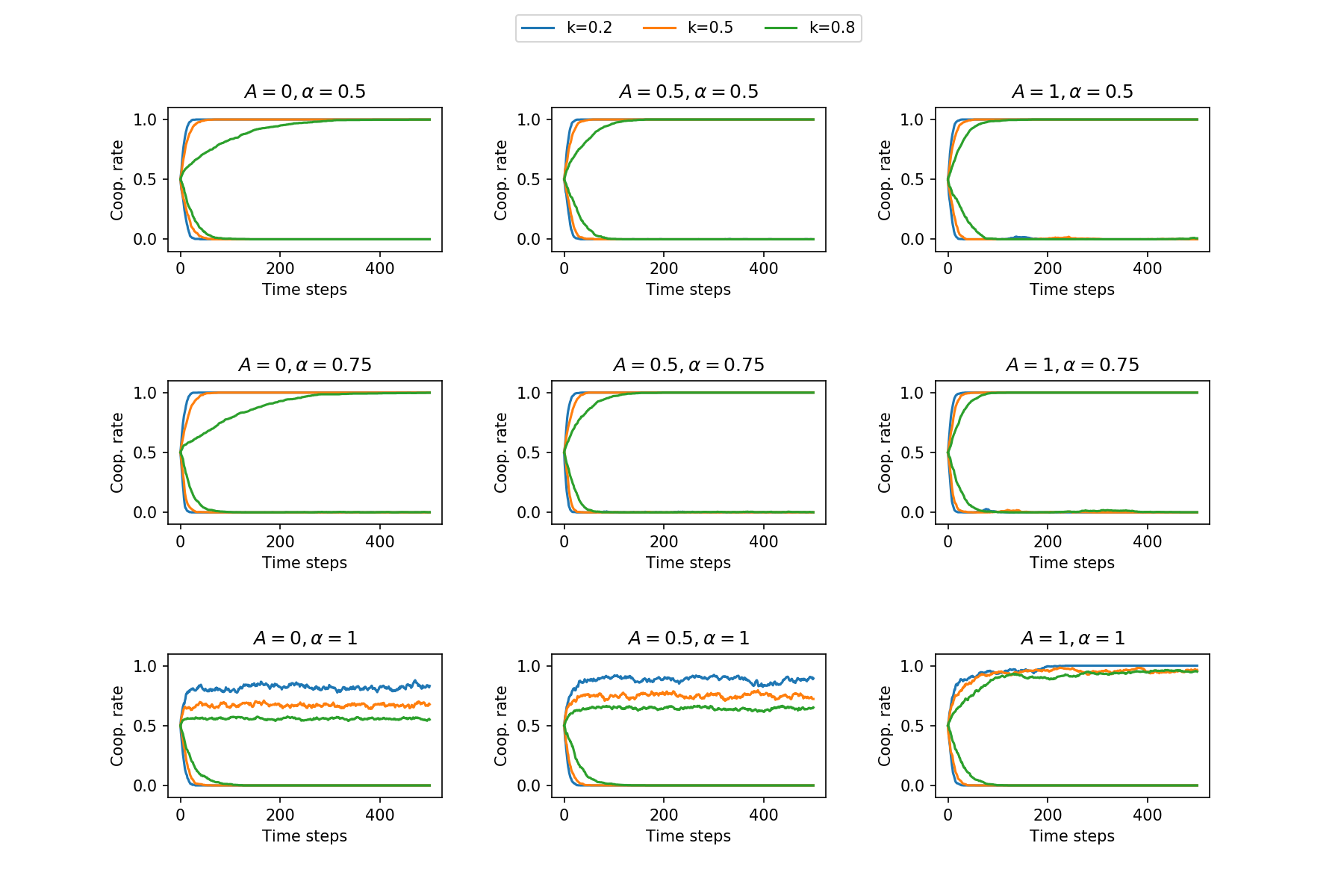}
    \caption{Evolution over time of the average rate of cooperation under deliberation when playing the one shot and the repeated prisoner dilemma for different values of the parameters $K$, $A$, and $\alpha$.}
    \label{fig:deliberation-Q-learning}
\end{figure}

In Figure \ref{fig:deliberation-Q-learning} we plot the evolution in time of the average rate of cooperation under deliberation when playing the one shot and the repeated prisoner dilemma for different values of the parameters $K$, $A$, and $\alpha$. The parameter $p$ is constant and equal to 0.5 to maintain a situation of symmetry between the two games. Colors are related to different values of $K$, as reported in the legend. In each subplot, for each color, there are two lines, one increasing and the other decreasing. The increasing line is the average cooperation rate under deliberation for the repeated prisoner dilemma, while the decreasing one is the average cooperation rate under deliberation for the one shot prisoner dilemma. 

When $\alpha$ is lower than one, i.e., equal to $0.5$ and $0.75$, agents quickly learn to cooperate under deliberation when they play the repeated prisoner dilemma, while they defect when playing the one shot interaction. In these cases simulations are qualitatively the same of the baseline model: after few iterations, agents cooperate with probability one when they deliberate in the repeated interaction, while they cooperate with probability zero when they deliberate in the one shot interaction. When $\alpha=1$, agents are able to learn to defect under deliberation in the one shot prisoner dilemma, while they are not able to completely learn to cooperate in the repeated prisoner dilemma under deliberation. This result is due to the fact that cooperation in the repeated interaction is the weakly dominant strategy, and not the strongly dominant one. Thus, agents who save in memory only the last payoff obtained, given $\alpha=1$, are often indifferent between the two actions and take their choice randomly. 

In general, we observe that the greater is $K$, the slower is the learning process under deliberation because it is less frequent. Furthermore, the learning process is quicker in the one shot interaction than in the repeated one, which is probably due the different types of dominance (strong as opposed to weak) in the two types of interaction. Moreover, higher assortativity and smaller learning rate speed up the learning process.

\section{The optimality of dual process reasoning}\label{append_optimal_k}
\begin{figure}[ht]
    \centering
    \includegraphics[width=0.7\linewidth]{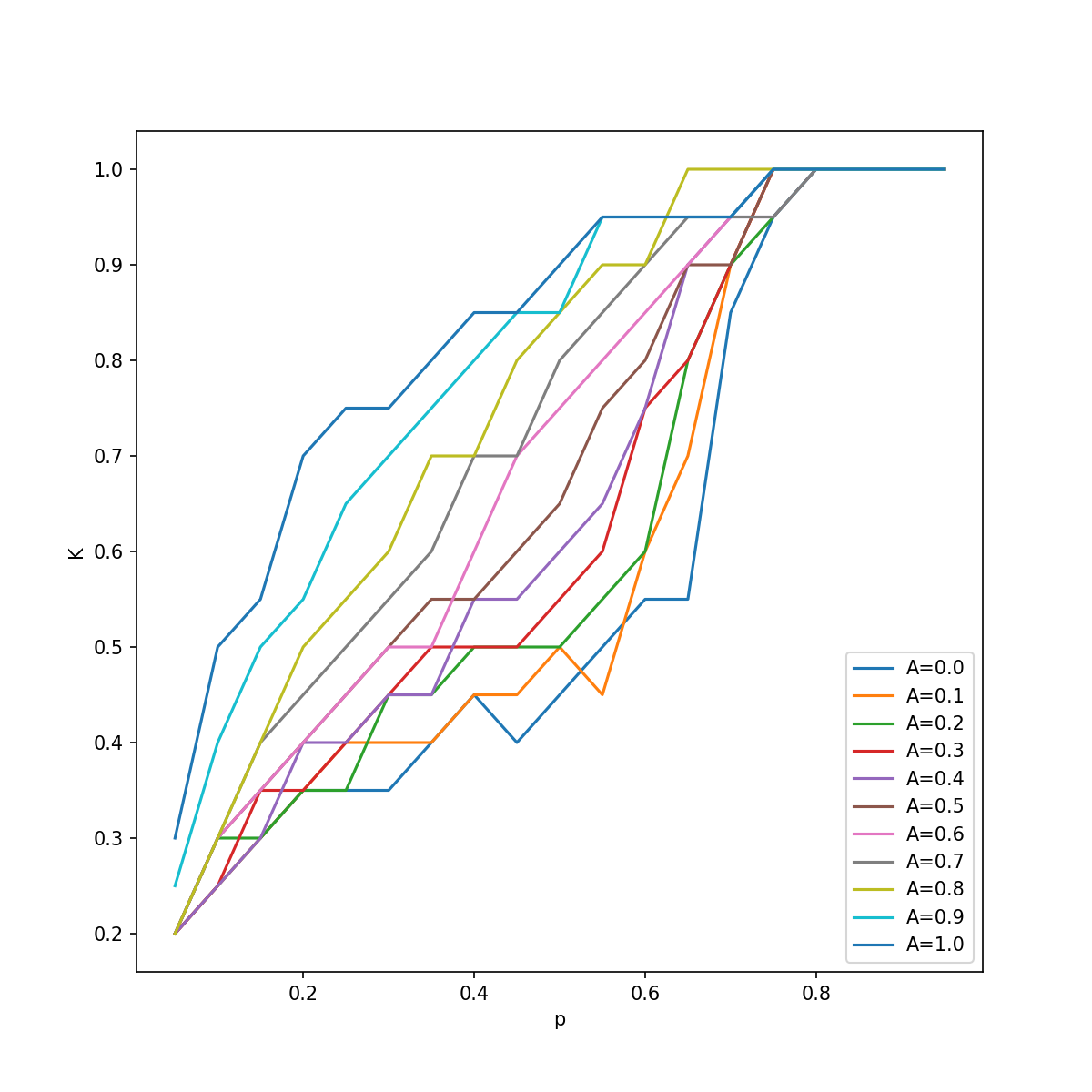}
    \caption{Values of $K$ that maximizes total cooperation varying the parameter $p$. Different lines with different colors are associated to different values of $A$.}
    \label{fig:Evolution-k}
\end{figure}
As we argue in the paper, our aim is not to study the evolution of dual process reasoning. Rather, we assume that agents are dual process reasoners, for which the literature has already provided evolutionary arguments, and we focus on the effects of assortativity in cognition on cooperation. Nevertheless, it can be interesting to notice that the value of $K$ that, for given $p$ and $A$, maximizes the overall level of cooperation is often strictly in between 0 and 1. This means that a population of dual process reasoners would perform better than a population of agents that are purely intuitive or deliberative agents.

A number of remarks can be done by looking at Figure \ref{fig:Evolution-k}. First, we notice that the value of $K$ that maximizes total cooperation increases as $p$ increases. In other words, a higher probability of repeated interactions, for which cooperation performs better than defection at the individual level, makes deliberation less important to maximize cooperation. Second, the higher the assortativity in cognition, the greater is the optimal value of $K$. Indeed, when assortativity is high, deliberation is more effective in shaping the heuristics, and thus less deliberation is needed to sustain cooperation. Third, we notice that populations of purely deliberative agents (i.e., $K=0$) are never optimal, while populations of purely intuitive agents (i.e., $K=1$) can be optimal, which happens when $p$ is high enough. These remarks appear not be fully general by looking at the figure (see, for instance, that the lines are not always monotonically increasing), but this can be an artifact of a rather limited number of simulations, also considering that we find a tiny difference of cooperation rates between the maximizing $K$ and the level of $K$ attaining the second-highest cooperation rate.


\bibliographystyle{unsrt}

\end{document}